\documentclass[prl,aps,preprintnumbers,twocolumn,showpacs, nofootinbib,superscriptaddress,notitlepage,10pt]{revtex4-1}
\usepackage{mathrsfs}
\usepackage{color}
\usepackage{enumitem} 
\usepackage{amsfonts}
\usepackage{amsmath}
\usepackage{slashed}
\usepackage{array}
\usepackage{verbatim}
\usepackage{epsfig}
\usepackage{graphicx}
\usepackage{color}

\newcommand{\beq}{\begin{eqnarray}}
\newcommand{\eeq}{\end{eqnarray}}

\newcommand{\nn}{\nonumber}

\begin{document}

\preprint{}

\title{Moments from Momentum Derivatives in Lattice QCD}
\affiliation{School of Science and Engineering, The Chinese University of Hong Kong, Shenzhen 518172, China}

\author{Zhuoyi Pang}
\affiliation{School of Science and Engineering, The Chinese University of Hong Kong, Shenzhen 518172, China}
\affiliation{University of Science and Technology of China, Hefei, Anhui, 230026, China}
\author{Jian-Hui Zhang}
\email{Corresponding author: zhangjianhui@cuhk.edu.cn}
\affiliation{School of Science and Engineering, The Chinese University of Hong Kong, Shenzhen 518172, China}

\author{Dian-Jun Zhao}
\email{Corresponding author: zhaodianjun@cuhk.edu.cn}
\affiliation{School of Science and Engineering, The Chinese University of Hong Kong, Shenzhen 518172, China}

%\date{\today}

\begin{abstract}
We show that the traditional moments approach in lattice QCD, based on operator product expansion (OPE), can be realized in a way that utilizes derivatives in momentum rather than in distance. This also avoids power divergent mixings, and thus allows to extract moments order by order, to all orders in principle. Moreover, by exploiting the symmetry of lattice matrix elements, we can determine the even and odd moments separately. As a demonstrative example, we determine the first three moments beyond the tensor charge $g_T$ of the isovector quark transversity distribution in the nucleon.

\end{abstract}
\maketitle

%\section{Introduction}
{\it \noindent Introduction:}
In high-energy scattering, the structure of hadrons is characterized by quantities such as parton distribution functions (PDFs), which describe the momentum distributions of quark and gluon partons inside a hadron. While these functions are essential for interpreting experimental data in hadron-hadron or lepton-hadron collisions, calculating them from first principles has been challenging due to their intrinsic nonperturbative nature and the fact that their definition involves lightcone correlations.  

Phenomenologically, PDFs are usually determined through global fits to a wide range of experimental data from high-energy collisions. As different fitting groups may choose different data sets and PDF parametrizations, such global fits can lead to ambiguities in the extracted PDFs, especially in certain kinematic regions where experimental data are sparse. On the other hand, lattice QCD offers a reliable first-principles approach that can provide important complementary information about PDFs. Traditionally, PDFs are accessed indirectly on lattice through the operator product expansion (OPE), where lightcone correlators are expanded in terms of local operator matrix elements that define the Mellin moments of PDFs. While these moments can be computed on the lattice, only the first few orders are typically obtainable due to potential power-divergent operator mixings at higher orders. Nevertheless, considerable efforts have been made to calculate the Mellin moments, and the results have had a valuable impact on phenomenological analyses (see, e.g., Ref.~\cite{Constantinou:2020hdm} for a recent review). A notable example is the global analysis of the quark transversity PDF in the nucleon~\cite{Lin:2017stx}, which incorporates the lattice result on the isovector tensor charge $g_T$ together with experimental data from semi-inclusive deep-inelastic scattering. The inclusion of the former has significantly reduced the uncertainty bands of the fit.

In addition to calculating moments, recent theoretical developments~\cite{Liu:1993cv,Detmold:2005gg,Braun:2007wv,Ji:2013dva,Ji:2014gla,Ma:2017pxb,Radyushkin:2017cyf,Chambers:2017dov} have enabled direct calculations of the Bjorken-$x$ dependence of PDFs from lattice. A commonly used quantity in these approaches is the equal-time Euclidean correlation function, which can be linked to PDFs either through a short distance factorization in coordinate space~\cite{Radyushkin:2017cyf} or through a large momentum factorization in momentum space~\cite{Ji:2013dva,Ji:2014gla,Ji:2020ect}, where the latter has been formulated as the large-momentum effective theory (LaMET)~\cite{Ji:2014gla,Ji:2020ect}. The LaMET approach has shown promise in determining the Bjorken-$x$ dependence of PDFs in the moderate $x$ region, while in the small and large $x$ regions the prediction is not reliable because power-suppressed higher-twist contributions become important there and must be taken into account. As a result, it is difficult to obtain moments in a reliable way from the integration of extracted PDFs over the momentum fraction $x$. In contrast, the short distance factorization approach relies on a global fit of moments to lattice matrix elements~\cite{Gao:2022iex,Gao:2022uhg,Gao:2023ktu}, which often used correlations at large distances where factorization is expected to fail. Furthermore, such a global fit suffers from similar ambiguities as those encountered in phenomenological fits to experimental data.

In this work, we show that, by leveraging recent theoretical developments, the traditional moments approach based on OPE can be realized
in a way that utilizes difference in momenta instead of distances. This also avoids power divergent mixings, and thus allows to extract moments order by order, to all orders in principle. Moreover, by exploiting the symmetry of lattice matrix elements, we can determine the even and odd moments separately. This approach avoids using long distance correlations  and focuses on relatively small momenta. As a result, it significantly reduces the computational cost of higher-order moments on the lattice, thereby greatly facilitating the reconstruction of PDFs across the full kinematic range from their Mellin moments and complementing ongoing efforts to extract PDFs from LaMET.

\vspace*{.5em}
{\it \noindent Traditional moments approach based on OPE:}
We briefly review the traditional approach for calculating moments on the lattice by taking the transversity structure function as an example. The local operators from OPE that measure the $(n-1)$-th moments $\left<x^{n-1}\right>$ take the following form
\beq\label{Eq:momop}
O_5^{\mu_1\mu_2\cdots\mu_n\perp}=\bar\psi(0)\gamma^{\{{\mu_1}}D^{\mu_2}\cdots D^{{\mu_n}\}}\gamma^{\perp}\gamma_5\psi(0),
\eeq
where an antisymmetrization between $\mu_1$ and $\perp$ indices is assumed, and the bracket indicates that the enclosed indices are symmetric and traceless. $\psi, \bar\psi$ denote the quark fields, and the covariant derivative
\beq
D_\mu&=\overleftrightarrow D_\mu=\frac{1}{2}(\overrightarrow D_\mu-\overleftarrow D_\mu)
\eeq
is realized on the lattice as following:
\begin{align}
\overrightarrow D_\mu\psi(x)&=\frac{1}{2a}\big[U_\mu(x)\psi(x+a\hat\mu)-U^\dagger_\mu(x-a\hat\mu)\psi(x-a\hat\mu)\big],\nn\\
\bar\psi(x)\overleftarrow D_\mu&=\frac{1}{2a}\big[\bar\psi(x+a\hat\mu)U^\dagger_\mu(x)-\bar\psi(x-a\hat\mu)U_\mu(x-a\hat\mu)\big],
\end{align}
where $U_\mu(x)$ is the link variable joining the points $x$ and $x+a\hat\mu$ with $\hat\mu$ being a unit vector.

While the operators in Eq.~(\ref{Eq:momop}) belong to irreducible representations of the Lorentz group in the continuum, they are in general reducible on the lattice and become linear combinations of irreducible representations of the hypercubic group, and therefore lead to operator mixings. In particular, for higher dimensional operators, the mixing coefficient can be power divergent, making the calculation of moments beyond the first few orders very challenging~\cite{Capitani:2002mp}.

\vspace*{.5em}
{\it \noindent Moments from nonlocal correlators:}
Now we outline our proposal in which we use difference in momenta rather than distances as a leverage to extract moments at a given order. The discussion below is based on the so-called quasi-light-front (quasi-LF) correlations, which are equal-time nonlocal quark and gluon bilinear operator matrix elements used in LaMET and short distance factorization. However, the same strategy can also be applied to other correlations used, e.g., in current-current correlations~\cite{Braun:2007wv} and lattice cross sections~\cite{Ma:2017pxb}, provided that a similar factorization formula exists. In the following, we take the isovector quark transversity PDF as an example. The discussion can be readily extended to other nonsinglet quark PDFs, as well as singlet quark and gluon PDFs, with minor modifications.

We consider the following quasi-LF correlation
\begin{equation}
\tilde h(z^2,\lambda=z P^z)=N\langle PS_\perp|\bar\psi(z)\gamma^t \gamma^{\perp}\gamma_5 W(z,0)\psi(0)|PS_\perp\rangle, 
\end{equation}
where $\psi, \bar\psi$ denote quark fields, $W(z,0)={\cal P}\exp[-ig\int_0^z du\, n\cdot A(u n)]$ is the gauge link along the $z$ direction with $n^\mu=(0,0,0,1)$, $|PS_\perp\rangle$ denotes an external hadron state with momentum $P^\mu=(P^t,0,0,P^z)$ and transverse polarization $S_\perp$, and $\lambda=z P^z$ is the so-called quasi-LF distance. %$N_\Gamma$ is a normalization factor with $N_{\gamma^t}=1/(2P^t), N_{\gamma^z\gamma_5}=1/(2P^z)$ and $N_{\gamma^t\gamma^\perp\gamma_5}=1/(2P^t S_\perp)$ for the unpolarized, helicity and transversity quark correlator, respectively. 
$N=1/(2P^t)$ is a normalization factor.
The nonlocal quark bilinear operator defining $\tilde h$ has been shown to renormalize multiplicatively~\cite{Ji:2017oey,Ishikawa:2017faj,Green:2017xeu}. Therefore, $\tilde h$ can be nonperturbatively renormalized by dividing by the same correlation in a zero-momentum hadron state
\beq
\tilde h_R(z^2,\lambda)=\frac{\tilde h(z^2,\lambda)}{\tilde h(z^2,\lambda=0)},
\eeq
which is known as the ratio scheme~\cite{Radyushkin:2017cyf}. As long as $z$ is within the perturbative region, this is a legitimate renormalization scheme, and has been used in many lattice calculations of PDFs. For recent examples, see, e.g.,~\cite{HadStruc:2021qdf,LatticeParton:2022xsd,Holligan:2023rex,Cloet:2024vbv}. The renormalized quasi-LF correlation at small $z$ can then be factorized into the LF correlation defining the leading-twist PDF, denoted as $h$, as following~\cite{Radyushkin:2017cyf,Zhang:2018ggy,Izubuchi:2018srq}
\begin{align}\label{eq:fac}
\tilde h_R(z^2, \lambda)&=\int_{-1}^1 du\, C(u, z^2\mu^2) \frac{h(u\lambda,\mu)}{h(\lambda=0,\mu)}+ p.c.,
%&=\int_{-1}^1 du\, C(u, z^2\mu^2)\int_{-1}^1 dx\,e^{-ix u\lambda}f_(x)\nn\\
\end{align}
where $\mu$ is the renormalization scale in the $\overline{\rm MS}$ scheme, $p.c.$ denotes power corrections. Using OPE, the above equation can be turned into an equation for the moments of PDFs~\cite{Izubuchi:2018srq,Wang:2019tgg}
\begin{align}\label{eq:momentsfac}
&\tilde h_R(z^2,\lambda)=N\sum_{n=1}^\infty \frac{(-iz)^{n-1}}{(n-1)!}C^{(n-1)}(z^2\mu^2)\nn\\
&\times\frac{\langle PS_\perp|n^t_{\mu_1} n_{\mu_2}...n_{\mu_n}O_5^{\mu_1...\mu_n\perp}(\mu)|PS_\perp\rangle}{g_T} + p.c.\nn\\
&=\sum_{n=1}^\infty \frac{(-iz P^z)^{n-1}}{(n-1)!}C^{(n-1)}(z^2\mu^2)\frac{\langle x^{n-1}\rangle}{g_T} + p.c.,
\end{align}
where
\begin{align} \label{Wil_moment}
\langle x^{n-1}\rangle&=\int_{-1}^1 dx\, x^{n-1}\delta q(x,\mu), \nn\\
C^{(n-1)}(z^2\mu^2)&=\int_{-1}^1 du\, u^{n-1}C(u, z^2\mu^2)
\end{align}
are the $(n-1)$-th moments of the isovector quark transversity PDF $\delta q(x,\mu)$ and the perturbative matching kernel $C(u, z^2\mu^2)$, respectively, $g_T=\int_{-1}^1 dx\,\delta q(x,\mu)$ is the tensor charge and $C^{(0)}=1$. The power correction $p.c.$ in Eq.~(\ref{eq:momentsfac}) consists of two parts: target mass corrections and genuine higher-twist contributions. Both have less powers in $P^z$ than the leading term as they arise from the trace terms. While the leading target mass corrections have been calculated to all orders for unpolarized, helicity and transversity quark PDFs in Ref.~\cite{Chen:2016utp}, genuine higher-twist contributions can be kept small by requiring $z^2\Lambda_{\rm QCD}^2\ll 1$.

Our proposal is based on the observation of different momentum dependence in moments of different orders, as well as in the leading and subleading power terms in the expansion above. Furthermore, the symmetry of real and imaginary parts of the quasi-LF correlation $\tilde h_R$ provides a practical advantage that greatly simplifies the calculation.

To show how to extract the moments order by order, we separate the real and imaginary parts of $\tilde h_R$ as
\begin{align}\label{eq:reim}
Re\, \tilde h_R&=\sum_{k=0}^\infty \frac{(-i z P^z)^{2k}}{(2k)!}C^{(2k)}(z^2\mu^2)\frac{\langle x^{2k}\rangle}{g_T} + p.c.,\\
Im\, \tilde h_R&=\sum_{k=0}^\infty \frac{(-i z P^z)^{2k+1}}{(2k+1)!}C^{(2k+1)}(z^2\mu^2)\frac{\langle x^{2k+1}\rangle}{g_T} + p.c..\nn
\end{align}
They are even and odd functions of $P^z$, respectively. Therefore, we can define a new variable $\zeta=(P^z)^2$, and the even and odd moments are then given by the following derivatives w.r.t. $\zeta$ at fixed $z$
\begin{align}\label{eq:even-odd-mom}
\frac{\langle x^{2k}\rangle}{g_T}&=\frac{(2k)!}{k!(-z^2)^k}\frac{1}{C^{(2k)}(z^2\mu^2)}\left.\frac{d}{d\zeta^k}Re\,\tilde h_R\right|_{\zeta=0}+p.c.,\nn\\
\frac{\langle x^{2k+1}\rangle}{g_T}&=\frac{(2k+1)!}{k!(-z^2)^k}\frac{1}{C^{(2k+1)}(z^2\mu^2)}\left.\frac{d}{d\zeta^k}\frac{Im\,\tilde h_R}{-i\lambda}\right|_{\zeta=0}+p.c.,
\end{align}
where $p.c.$ denotes ${\cal O}(z^2M^2, z^2\Lambda_{\rm QCD}^2)$ corrections that can be kept small by keeping the distance $z\ll 1/\Lambda_{\rm QCD}$. In addition, these power corrections will be further suppressed by extra combinatorial factors at each order, so that they only have a minor impact on the extraction of moments at a given order. One great advantage of using Eq.~(\ref{eq:even-odd-mom}) is that by using symmetry of the renormalized matrix elements and defining the new variable $\zeta$, the even and odd moments can be determined separately, therefore it is much easier to go to higher orders. 

Following the same spirit, the even moments can also be extracted with a reduced number of derivatives:
\begin{equation} \label{eq:new_even}
    \frac{\big<x^{2+2k}\big>}{g_T}=\frac{(2+2k)!}{k!(-z^{2})^{1+k}}\frac{1}{C^{(2+2k)}(\mu^{2}z^{2})}\frac{d\frac{\text{Re}\tilde{h}_{R}-1}{\zeta}}{d^{k}\zeta}\Big|_{\zeta=0}+p.c.,
\end{equation}
where $k\geq0$. A special case is the extraction of the second moment, which is free from taking derivatives:
\begin{equation}
    \frac{\big<x^{2}\big>}{g_T}=\frac{2}{-z^{2}}\frac{1}{C^{(2)}(\mu^{2}z^{2})}\frac{\text{Re}\tilde{h}_{R}-1}{\zeta}\Big|_{\zeta=0}+p.c..
\end{equation}
We use the combination $\text{Re}\tilde{h}_{R}-1$ in \eqref{eq:new_even} to subtract the mass corrections of $\mathcal{O}({\zeta^0})$ in $\text{Re}\tilde{h}_{R}$ following Ref.~\cite{Chen:2016utp}, so that $d\frac{\text{Re}\tilde{h}_{R}-1}{\zeta}/d^{k}\zeta$ is regular at $\zeta=0$. 

\vspace*{.5em}
{\it \noindent Numerical tests:} Using the approach above, we extract the first three moments beyond $g_T$ of the isovector quark transversity PDF in the nucleon, based on the {renormalized matrix elements} at a single lattice spacing $a=0.094$ fm with a pion mass 358 MeV and various momenta in Ref.~\cite{HadStruc:2021qdf}. All moments presented here have been normalized by $g_T$. The next-to-leading order Wilson coefficients needed in Eq.~(\ref{eq:reim}) have also been given in Ref.~\cite{HadStruc:2021qdf} as
\begin{align}
C^{(n)}(z^2\mu^2)&=1+\frac{\alpha_s C_F}{\pi}\Big[\ln\frac{z^2\mu^2 e^{2\gamma_E+1}}{4}\sum_{k=2}^{n+1}\frac 1 k\nn\\
&-\Big(\sum_{k=1}^n \frac 1 k\Big)^2-\sum_{k=1}^n\frac{1}{k^2}\Big],
\end{align}
where $\alpha_s$ is the strong coupling constant, $\gamma_E$ is the Euler constant.

\begin{figure}[thbp]
\includegraphics[width=.5\textwidth]{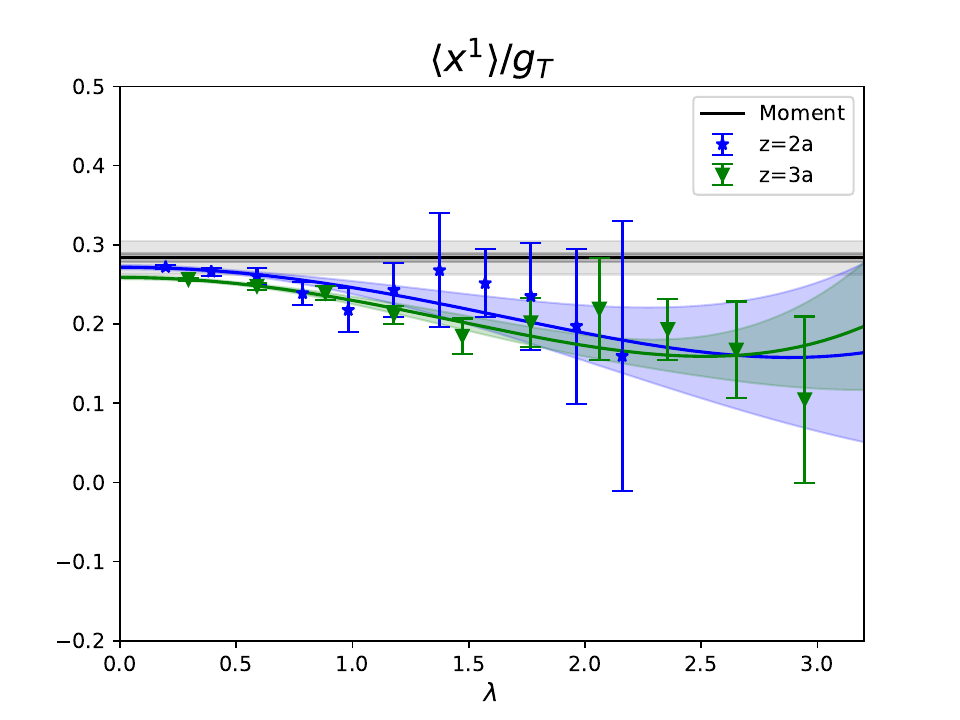}
\includegraphics[width=.5\textwidth]{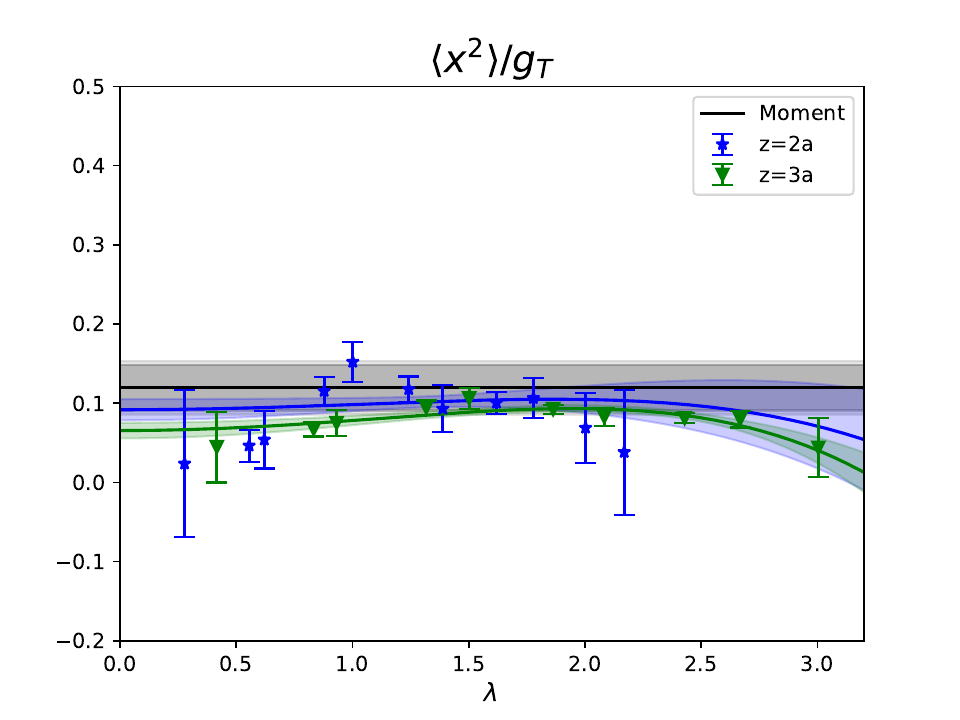}
\includegraphics[width=.5\textwidth]{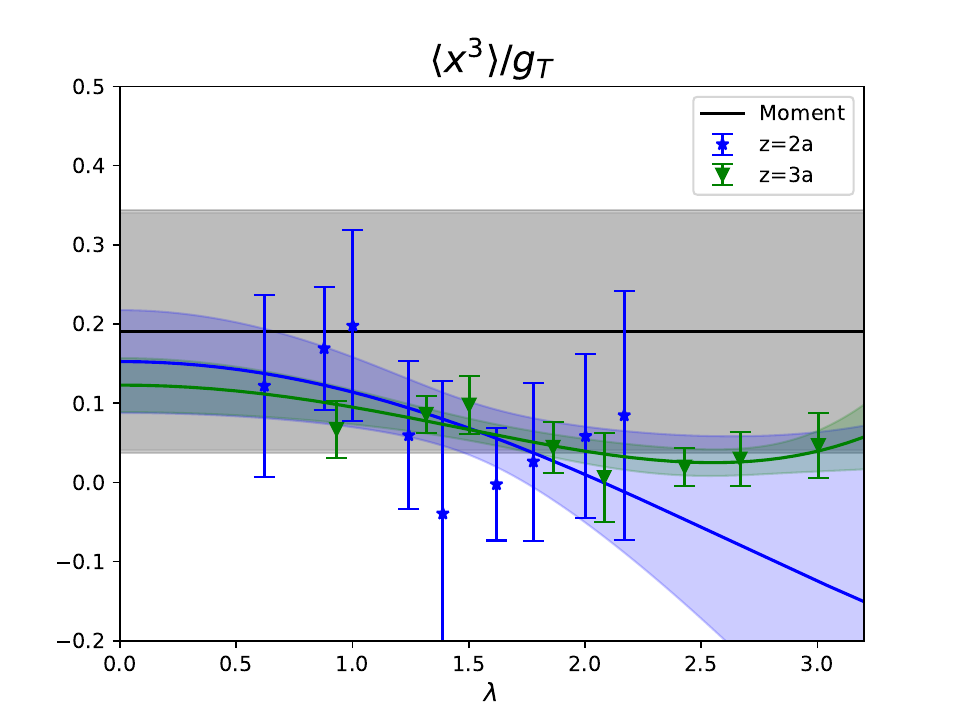}
\caption{The first three moments beyond the tensor charge $g_T$ of the isovector quark transversity PDF in the nucleon. All results are normalized by $g_T$. The renormalization scale is $\mu=\sqrt 2$ GeV. The blue stars and green inverted triangles represent the raw data for joint fit lines of $z=2a$ and $z=3a$, respectively, while the black line and gray band show the results of $\langle x^n\rangle/g_T$.} 
\label{fig:moments}
\end{figure} 

Note that in our approach, the derivatives of lattice matrix elements are taken w.r.t. the momentum square $\zeta$. Data at different perturbative distances, $z$, can be combined together to determine moments at a given order. This is also the strategy we adopt here due to the limited data quality. {Taking the extraction of $\langle x^{2}\rangle/g_{T}$ as an example, we start from the following formula based on OPE:}
\begin{align} \label{joint}
\frac{-2}{z^2C^{(2)}}\frac{d Re\tilde{h}_R}{d\zeta}=&\frac{\langle x^2 \rangle}{g_T}(\mu) \nonumber \\ 
&+\lambda^2\frac{C^{(4)}}{C^{(2)}}b(\mu)+\lambda^4\frac{C^{(6)}}{C^{(2)}}c(\mu) \nonumber \\
&+z^2\Lambda_{\text{QCD}}^2\frac{d(\mu)}{C^{(2)}}+z^2\Lambda_{\text{QCD}}^2\lambda^2\frac{e(\mu)}{C^{(2)}},
\end{align}
{where we have truncated the expansion up to terms of $O(\lambda^4)$ in the leading-twist OPE and included leading higher-twist contributions as well. $C^{(n)}$s are defined by Eq.~\eqref{Wil_moment}, their dependence on $\mu^{2}z^{2}$ is not shown for simplicity. To extract $\langle x^{2}\rangle/g_{T}$, an extrapolation to $\lambda=0$ (or equivalently, to $P_z=0$) needs to be carried out, similar to the continuum extrapolation to $a=0$. We have tested that including higher-order terms in Eq.~\eqref{joint} only has a minor impact on the fitting result. %The second line denotes the third and fourth terms in the OPE of $Re\tilde{h}_{R}$ (Eq.(9)), while the third line denotes genuine higher-twist contributions. 
Besides $\langle x^{2}\rangle/g_{T}$, $b,c,d,e$ in the above equation are also free fitting parameters depending on the scale $\mu$, whose variation provides an important source of systematic errors. For extractions of $\langle x^{1}\rangle/g_{T}$ and $\langle x^{3}\rangle/g_{T}$, we use formulas similar to Eq.~\eqref{joint}, including the first three (two) terms at twist-2 (twist-4) level.}  
%the fitter learn the correct pattern expected from OPE··· insensitive to the prior. Actually, one should use small $pz$(such that $\lambda<1$) such that the higher-order is automatically negligible.

We have chosen lattice data at $z=2a ,3a$ for our analysis, as $z=a$ data suffer from discretization effects while $z\ge4a$ is too large so that the validity of factorization in Eq.~(\ref{eq:fac}) becomes questionable~\citep{Su:2022fiu}. We take the difference between including $z=a$ data or not as part of our systematic uncertainties. Our results are shown in Fig.~\ref{fig:moments} with %All moments have been normalized by the tensor charge $g_T$ with
\begin{align} \label{fit}
%\langle x^1\rangle/g_T&=0.277 (13)(24) ,\nn\\
\langle x^1\rangle/g_T&=0.283 (06)(20) ,\nn\\
%\langle x^2\rangle/g_T&=0.099 (25)(20) ,\nn\\
\langle x^2\rangle/g_T&=0.120 (29)(18) ,\nn\\
%\langle x^3\rangle/g_T&=0.17 (14)(01) ,
\langle x^3\rangle/g_T&=0.19 (15)(01) ,
\end{align}
where the renormalization scale has been chosen as $\mu=\sqrt 2$ GeV, and the numbers in the two parentheses represent statistical and systematic errors, respectively. The sources of systematic errors are mainly composed of two parts: 1) Inclusion of data at $z=a$: The lattice data at $z=a$ have large discretization errors. We consider the difference in our results with and without including these data as part of our systematic uncertainties; 2) Perturbative running: We change the renormalization scale to $\mu=2$ GeV and include the difference in the results as systematic uncertainties. This scale dependence reflects the impact of missing higher-order perturbative corrections. 
Our final results for $\langle x^1\rangle$ and $\langle x^2\rangle$ are slightly larger than, but are still consistent within $1\sim 2 \sigma$ with those extracted in Ref.~\cite{HadStruc:2021qdf} through global fits. 
{In Ref.~\citep{HadStruc:2021qdf}, the transversity PDF was fitted (at $\mu=\sqrt{2}$ \text{GeV}) based on the OPE of $\tilde{h}_{R}$ using lattice data ranging from from $z=0.188$ fm to $z=0.94$ fm. Then the first and second moments beyond $g_T$ were obtained through the fitted transversity PDF. %The data points with a wide range of $z$ (from $z=0.188$ fm to $z=0.94$ fm) were used for the fit, resulting in 
This gives slightly different numbers and the appearance of smaller uncertainties in their results.
However, when $z\gtrsim 0.2$ fm, the errors of Wilson coefficients become large under scale variation, and the OPE shall not be trusted when $z\gtrsim 0.4$ fm.}
Our $\langle x^3\rangle$ has relatively large errors due to the limited data points and quality, which can be improved in the future by increasing statistics and collecting more data points at small momenta. Note that our approach does not require using lattice data at large distances, as done in~\cite{HadStruc:2021qdf,Gao:2022iex,Gao:2022uhg,Gao:2023ktu}, and our results do not suffer from ambiguities of global fits.

\vspace*{.5em}
{\it \noindent Prospect for lattice calculations:}
In contrast to the traditional moments approach, our method offers several advantages. First, by leveraging the multiplicative renormalizability of equal-time nonlocal bilinear operators, the potential power divergent mixing between higher- and lower-order moments operators has been avoided. Second, the usual differentiation w.r.t. distance has been replaced by a differentiation w.r.t. momentum, which is computationally much cheaper. Third, by using symmetry of the lattice matrix elements, the even and odd moments can be computed separately, making it much easier to go to higher orders. Fourth, there is no need to use long distance correlations where factorization is expected to fail, and the determination of higher-order moments is independent of the lower-order ones. 

To facilitate lattice calculations using the method presented here, we need lattices with a large box size $L$ so that the step in momentum $\Delta P=2\pi/L$ is small {and we 
shall focus on the small $\lambda$ region($\lambda\lesssim1$)}. We
can reliably take the derivatives and extrapolate to zero momentum, whereas a small lattice spacing is less important. In practical calculations, we can combine matrix elements at different perturbative distances to determine moments at a given order. 

\vspace*{.5em}
{\it \noindent Universality class of correlators:}
In the discussion above, we have focused on quasi-LF correlators employed in LaMET and short-distance factorization. However, the methodology we outlined can be applied to other types of nonlocal correlators, such as current-current correlations and lattice cross sections, where a factorization similar to Eq.~(\ref{eq:fac}) holds. These nonlocal correlators form a universality class of operators, whose matrix elements can be combined together to compute the same moments effectively. 

\vspace*{.5em}
%\section{conclusion}\label{concl}
{\it \noindent Conclusion and outlook:}
To conclude, we have shown that the traditional moments approach based on OPE can be realized in a way that utilizes derivatives w.r.t. momentum rather than distance. This is computationally much cheaper and avoids power divergent mixings, thus allows for moments calculations to be performed order by order, to all orders in principle. Taking the isovector quark transversity PDF of the nucleon as an example, we illustrated how its first three moments beyond $g_T$ can be computed through differentiating renormalized quasi-LF correlations w.r.t. momentum square. Furthermore, we showed how the real and imaginary parts of the correlations can be used to separately extract even and odd moments. We also outlined the main advantages and requirements for the lattice setup in our proposal. It will be intriguing to explore the extent to which we can extend the computation of moments to higher orders using the proposed approach. The results will provide valuable complementary insights into the partonic structure of hadrons along with other approaches.  

{\it \noindent Note added:} After this paper is submitted, we noticed that similar strategies {(taking differentiation of $\tilde{h}_R$ w.r.t. $\lambda$)} have also been discussed in Ref.~\cite{Karpie:2018zaz}, {whereas our proposal to take differentiation w.r.t. $\zeta=(P^z)^2$ rather than $\lambda=z P^z$ allows for a faster convergence in the extrapolation to zero momentum and a more efficient determination of higher moments. 
%For future calculations with smaller momentum steps, 
Moreover, the authors of Ref.~\cite{Karpie:2018zaz} discussed the extracted moments as a function of $z^2$, ranging from $z=a$ to nonperturbative $z$.
In contrast, we only use data points at $z=2a$ and $z=3a$ for our joint fit, which
avoids heavy contaminations from discretization errors and higher-twist effects.
We leave a more detailed analysis (with resummed Wilson coefficients) using data points generated by lattices with a larger box size in the future.}
%We leave a detailed analysis, including resummation dependence in a future work.
%perturbative $z$, 
%and thus avoid heavy contaminations from discretization errors and higher-twist effects.}
%due to the discrete error(invalidity of perturbative factorization) in such a region.}
We also discussed the universality class of operators in our paper.

\vspace*{.5em}
%\section{Acknowledgments}
\begin{acknowledgments}
%{\bf Acknowledgment:}
We are grateful to Jiunn-Wei Chen for valuable discussions at the early stage of this project. We also thank Joe Karpie for sharing their lattice data with us, and Huey-Wen Lin and Yi-Bo Yang for helpful discussions. This work is supported in part  by  National  Natural Science Foundation of China under Grants No. 12375080, 11975051, 12061131006, the Ministry of Science and Technology of China under Grant No. 2024YFA1611004, and by CUHK-Shenzhen under grant No. UDF01002851.
\end{acknowledgments}

%\bibliographystyle{apsrev4-1}
%\bibliography{bibliography}
\end{document}